\documentstyle[aps]{revtex}

\input{epsf}

\input{epsf}

\begin{document}



\title{MULTIPLE PARTON INTERACTIONS IN HIGH-DENSITY QCD MATTER}

\author{Dinesh K. Srivastava$^1$ and Klaus Geiger$^2$}

\address{$^1$ Variable Energy Cyclotron Centre, 1/AF Bidhan Nagar, Calcutta 700 064, India}
\address{$^2$ Physics Department, Brookhaven National Laboratory, Upton, N. Y. 11973, U. S. A.}

\date{\today}

\maketitle

\vspace{5.0cm}

\centerline{{\bf Abstract}}
\begin{abstract}
Multiple interactions of quarks and gluons in high-energy
heavy-ion collisions may give rise to interesting
phemomena of color charges propagating in high-density
QCD matter.
We study the dynamics of multi-parton systems produced
in nucleus-nucleus collisions at energies corresponding
the the CERN SPS and the future BNL RHIC experiments.
Due to the complexity of the multi-particle dynamics 
we choose to employ the parton cascade model in order to 
simulate the development of multiple parton scatterings
and associated stimulated emision processes.
Our results indicate a {\it non-linear} increase with
nuclear mass $A$ of, e.g., parton multiplicity, energy
density, strangeness, and contrasts
a {\it linear} $A$-scaling as in Glauber-type approaches.
If multiple interactions are suppressed and only
single parton scatterings (no re-interactions) are considered, 
we recover such a linear behavior.
It remains to be studied whether these results on the parton level
can be experimentally seen in final-state observables,
such as the charged particle multiplicity, the 
magnitude of produced transverse energy, or the number of produced
strange hadrons.
\end{abstract}

\vspace{5.0cm}

\leftline{e-mail: dks@vecdec.veccal.ernet.in, klaus@bnl.gov}

\vspace{1.0cm}

\leftline{Pacs No.: 12.38.Bx, 12.38.Mh, 25.75.+r, 24.85.+p}

\newpage

\section{INTRODUCTION}

The propagation of high-energy particles through matter
is generally subject to enhanced interaction rates as compared to free space,
and hence imply a modification of the propagators and vertices.
It is therefore of fundamental interest for studying
quantum field theory in a finite-density or finite-temperature environment.
An enhanced probability for multiple particle interactions 
can of course only arise if the distribution of potential interaction 
partners in the system is sufficiently dense.
Vice versa, the 
manifestation of multiple interaction effects in observable quantities
should reflect the structure of dense matter environment.
Hence, multiple interactions expose the dual relation between the
matter as a macroscopic medium, and the particles within it and 
generating its dynamics on the microscopic level:
\begin{itemize}
\item[a)]
If systematically studied, multiple interactions may
be used as a "densometer", i.e. a probe of the particle density
of the medium.
\item[a)]
Measuring characteristics of particles traversing
dense matter by encountering multiple interactions, can be used
to learn about the influence of the medium on the particle dynamics.
\end{itemize}
\smallskip

Multiple scattering phenomena in many-body systems 
has been widely studied in a variety of applications \cite{scott},
for example, electron showers in cosmic rays, 
electron scattering in crystals or emulsion,
or hadrons propagating through nuclear matter in collisions 
involving heavy nuclei.
In many applications, however, it has been treated as sequential
re-interactions of particles by means of {\sl elastic} scatterings off the 
targets in a medium. In this case  the particles essentially undergo
a simple random walk in phase-space that causes both momentum and spatial 
broadening of the particle spectrum.
The situation is very different, if one allows for
{\sl inelastic} interactions, in which case the particles not
only get deflected, but also transfer or absorb energy-momentum to or from the medium, 
and moreover can multiplicate themselves through excitation 
and decay that is stimulated by the energy transfer.
\medskip

In this paper we are concerned specifically with 
multiple interactions of quarks and gluons in a high-density
QCD matter environment that is created in 
ultra-relativistic heavy-ion collisions.
In these collisions short range parton interactions can play
a major role for the nuclear dynamics 
at least during the early and most dissipative stage of the first few 
$fm/c$ \cite{msrep}.
A copious production of 
quarks and gluons, liberated from the mother nuclei through 
(semi)hard parton-parton scatterings, is expected to initiate a 
complex evolution of internetted parton cascades that consist
of multiple parton (re)scatterings plus associated emission and absorption of gluons.
\medskip

As we are dealing with
quantum field theory, here QCD, a classical scattering picture is
certainly inappropriate. This is because the mere existence of 
particle fluctuations,
being created and annihilated, requires to account for scattering
processes (elastic and inelastic)
as well as  emission/absorption processes (virtual and real).
In particular, the interplay between collisions and stimulated emsissions is
 non-trivially entwined, since an increased number of scatterings can
initiate an enhanced particle multiplication through emissions (decays),
which in turn increases the density of particles and hence potential scattering targets
for further collisions, and that may stimulate further emissions.
Clearly, this avalanche effect cannot grow unbriddled, since unitarity
requires the cross-section to be finite (the medium cannot be blacker than black), 
and moreover screening effects will become the more important, the larger the density.
For these reasons, a `classical' cascade picture is clearly {\it not} applicable.
Fig. 1 illustrates schematically the differences between  `classical' cascades
and  `quantum' cascades.
\smallskip

It is evident that an exact treatment of multiple QCD interactions in a high-density
environment is a problem of (at present) intractable complexity, both
on the perturbative and the non-perturbative level.
It requires a detailed understanding of, e.g., the non-abelian interference
of interaction amplitudes, the propagation and screening of color charges
due to these interactions, the infra-red regime of very low energy/momentum
transfer, the confinment dynamics and hadron formation in dense matter.
Even though recent remarkable progress within the isolated issue of
parton propagation in finite-density matter has been made on the 
basis of QCD's first principles \cite{gyulassywang,BDMPS,qiu}, 
the results, being based
on highly idealized assumptions, are at present not useful for
experimental study and verification.
Therefore, in order to take pragmatical steps, we need to rely  on a model
decription that serves as a baby-version of the (impossible) exact 
QCD treatment.
\smallskip

\begin{figure}
\epsfxsize=350pt
\centerline{ \epsfbox{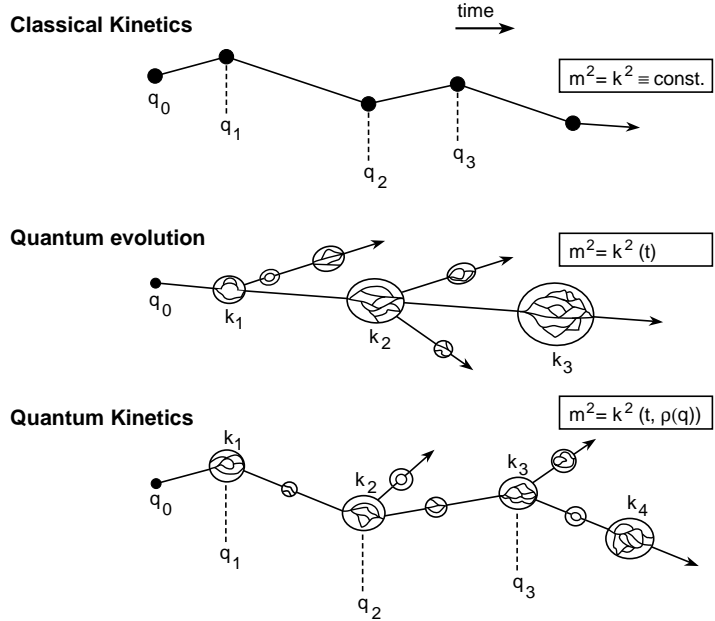} }
\caption{
(a) A `classical' cascade in a medium
as s succession of elastic 
(billiard-ball) scattering of stable, pointlike  particles with 
$k^2 = E^2-{\bf k}^2 = const$. 
(b) 
A `brems-strahlung' cascade in free space on the basis of
the renormalization-group evolution of unstable quanta with
time-dependent $k^2 =  k^2(t)$. 
(c) 
A `quantum' cascade in a medium by means of multiple scatterings with stimulated
real and virtual emission of quanta that have 
time- and density-dependent $k^2 =  k^2(t,\rho)$. 
}
\end{figure}

Specifically, we employ here 
the {\it Parton Cascade Model} (PCM) \cite{pcm},
to quantify the above intuitive picture of multiple interaction physics
in a phenomenological manner that makes contact with the
experiments at the CERN SPS and the BNL RHIC.
The PCM assumes
that the early stage of nuclear collisions at
sufficiently high energies can well be described in terms of the
space-time evolution of internetted parton cascades \cite{msrep},
based on renormalization-group improved perturbative QCD in conjunction with
quantum kinetic theory \cite{ms3942}, whereas the later non-perturbative development
of the transformation of partons to final-state hadronic states
must be treated within a phenomenological hadronization scheme.
Hence, within this  framework, the early stage of parton evolution
rests on the robust principles of  perturbative QCD,
whereas the modeling of the later development of parton-hadron conversion 
has no such firm theroretical basis, and moreover is not unique
in view of lack of knowledge of the underlying confinement mechanism.
Therefore we focus our analysis on the partonic sector, which is free of
the ambiguities of a particular hadronization model. However, we 
plan to investigate to which degree multiple parton interactions
may reflect themselves in final-state hadron yields. This latter aspect
requires a parton-hadron conversion model, for instance 
the hadronization scheme of Ref. \cite{EG} in which
partons coalesce to colorless clusters as controlled by the local density
of color charges and the subsequent decay of these prehadronic clusters
into hadron states.
In the PCM, the colliding nuclei are 
viewed as two coherent clouds of valence quarks plus
virtual gluon- and seaquark-fluctuations 
that materialize into ``real'' excitations due to primary
parton-parton scatterings.  For large nuclei, this primary parton production
 can result in a large initial particle and energy density in the
central collision region, which increases further by subsequent intense
gluon bremsstrahlung, secondary scatterings and rescatterings.
Detailed balance is assured by including absorption or fusion
of partons when the local density becomes so large that
the quanta begin to overlap in phase-space, so that the
unitarity principle is obeyed.
\medskip

In the remainder of the paper, we report and discuss results
obtained with the Monte Carlo program VNI \cite{vni}
which allows one to simulate heavy-ion collisions on the basis of the
aforementioned PCM concepts.
Within this model framework, we address 
effects of multiple parton scattering
and associated multi-gluon radiation during nucleus-nucleus
collisions at the highest beam energies, corresponding to
$\sqrt{s} \simeq 20$ A GeV at the CERN SPS accelerator 
and $\sqrt{s} = 200$ A GeV at the BNL RHIC setup.
 From a phenomenological point of view it is important
to investigate whether
the final-state particles carry some information 
about their partonic interaction history into the detectors, and whether
it is possible to extract observable manifestations of
multiple parton interactions in the final-state spectra.
Although we do not address these connections to the final hadronic
yields and spectra (which we postpone to a sequel to this work) here,
we think that the partonic dynamics serves as an estimate of the
type and magnitude of observable effects.
\smallskip

Specifically, we focus on three selected topics, which, we believe, 
are of foremost interest:
\begin{itemize}
\item
{\it Development of initial conditions:}
What is the characteristic development of the initial
partonic system during the first fraction of a $fm/c$?
How fast does the materialization of partons through
collisions and emissions occur?
How large is the resulting particle and energy density of the
partonic system that determines the initial conditions for the
subsequent expansion, hadronization and freeze-out?
\item
{\it Multiple scattering as response to dense matter:}
What are the significances and potential experimentally
measurable implications
of multiple parton interactions as a response to the high-density matter?
What is the growth pattern of the total partonic entropy and transverse energy
when  the size of the nuclei is increased? Do multiple scatterings
result in a visible non-linear growth with 
nuclear mass number $A$, in contrast to a linear scaling as in
Glauber models?
\item
{\it Interplay between multiple scatterings and parton emissions:}
How large  is the effect 
(e.g., on rapidity density and transverse energy)
of multiple scatterings compared
to single scatterings of partons? 
What is the relative importance of multiple parton collisions
and associated parton emissions? 
How different are the parton spectra with/without
multiple interactions and with/without stimulated emissions? 
\end{itemize}
\smallskip

\section{Multiple parton interactions during the early stage
of heavy-ion collisions at CERN SPS and RHIC energies}
\medskip

Since the notion of multiple interactions is {\it defined} as successions of
similar processes that change the direction of momentum
(and possibly the invariant mass) and which are statistically
independent (at least approximately), it is clear that truly quantum-mechanical
interference between the interaction amplitudes cannot be accounted for
$-$ except when it is possible to describe them effectively
in a probabilistic manner. 
Moreover, in view of lack of knowledge of the details of new phenomena
associated with ``dense-matter QCD", we do not attempt to model unknown
medium effects, but rather restrict ourselves to those effects
that can arise from statistical many-body kinetics.
The main assumptions and approximations of the probabilistic 
parton cascade picture that we employ, are:
\begin{description}
\item[(i)]
Factorization of short distance interactions of cascading partons
with the nuclear medium from long range non-perturbative forces is
assumed.
This means, even in the presence of dense medium where a parton
can encounter multiple successive interactions, a probabilistic
description of local, non-interfering interactions applies at
sufficiently high energies.
\item[(ii)]
Non-trivial interference effects are generally neglected, {\it except}:
destructive interference of interaction
amplitudes in both, coherent succesive small angle
emissions (`angular orderered emission'), and multiple sequential scatterings
(`Landau-Pomerantchuk-Migdal effect').
\item[(iii)]
Nuclear effects due to long range correlations, e.g. nuclear
shadowing (anti-shadowing) which are not associated with truly perturbative
parton interactions are ignored.
Similarly, possible mean field effects aring from a collective
motion of the particle system are beyond our concern.
\end{description}
\medskip

In the following we first study the evolution of parton distributions
in central collision of two gold nuclei, at $\sqrt{s}\,=$ 20 A GeV and
200 A GeV. We first address the early stage of the
parton cascade development.
Then we turn to our results for the energy densities achieved in
collisions involving  several identical nuclei across the periodic table.
Finally we discuss characteristic effects
of multiple parton interactions as a response to the high-density matter.
\medskip

In our analysis of parton spectra and density profiles,
we have considered - aside from the initially present
valence quarks -  only {\it materialized} quarks and gluons that
have suffered {\it at least one interaction}, i.e.,
any parton that was either knocked out
from the nuclear wavefunction by a collision with another
initial-state parton (primary scattering) or
with an already interacted parton (rescattering), 
or being  created through a pair production, or emitted by
a gluon bremsstrahlung process.
Moreover,
we have considered only (semi)hard parton-parton collisions
with momentum transfer $q^2 \ge p_0^2$, where $p_0 = 1.1$ GeV/$c$ for 
$\sqrt{s} = 20$ A GeV and $p_0 = 2.1$ GeV/$c$ for $\sqrt{s} = 200$ A GeV.
These values for $p_0$ correspond to the default values \cite{vni} of
VNI, which were found to agree well with measured particle spectra
from $pp$ collisions at ISR to Tevatron energies, as well as from
$AB$ collisions at CERN SPS energy \cite{dkskkg1}.
In principle one could extend to include parton scatterings
with $q^2 < p_0^2$, as discussed in Refs. \cite{msrep,vni},
by allowing for the whole range $0 \le q^2 \le (p_1+p_2)^2$. 
In this case one would divide (semi)hard 
($q^2 \ge p_0^2$) and soft ($q^2 < p_0^2$)
parton collisions and matching them at $p_0^2$,
the former treated perturbatively and the latter described by a 
reasonable phenomenological ansatz.
However, in order to avoid the ambiguities and model-dependence
associated with non-perturbative  soft scatterings, we choose
to restrict ourselves to include exclusively (semi)hard scatterings for
which we have a confident, perturbative description available.
\medskip

\subsection{Development of initial conditions}
\smallskip

We start by comparing the parton dynamics at
$\sqrt{s} =20$ A GeV and $\sqrt{s}= 200$ A GeV during the 
first 0.5 $fm/c$ after nuclear overlap,   the stage that 
determines the initial conditions for the subsequent expansion.
Figs. 2 - 5 sum up our findings regarding the
early time evolution of the parton distributions, starting at $t=0$ when
the nuclei overlap. 
\smallskip

\begin{figure}
\epsfxsize=250pt
\centerline{ \epsfbox{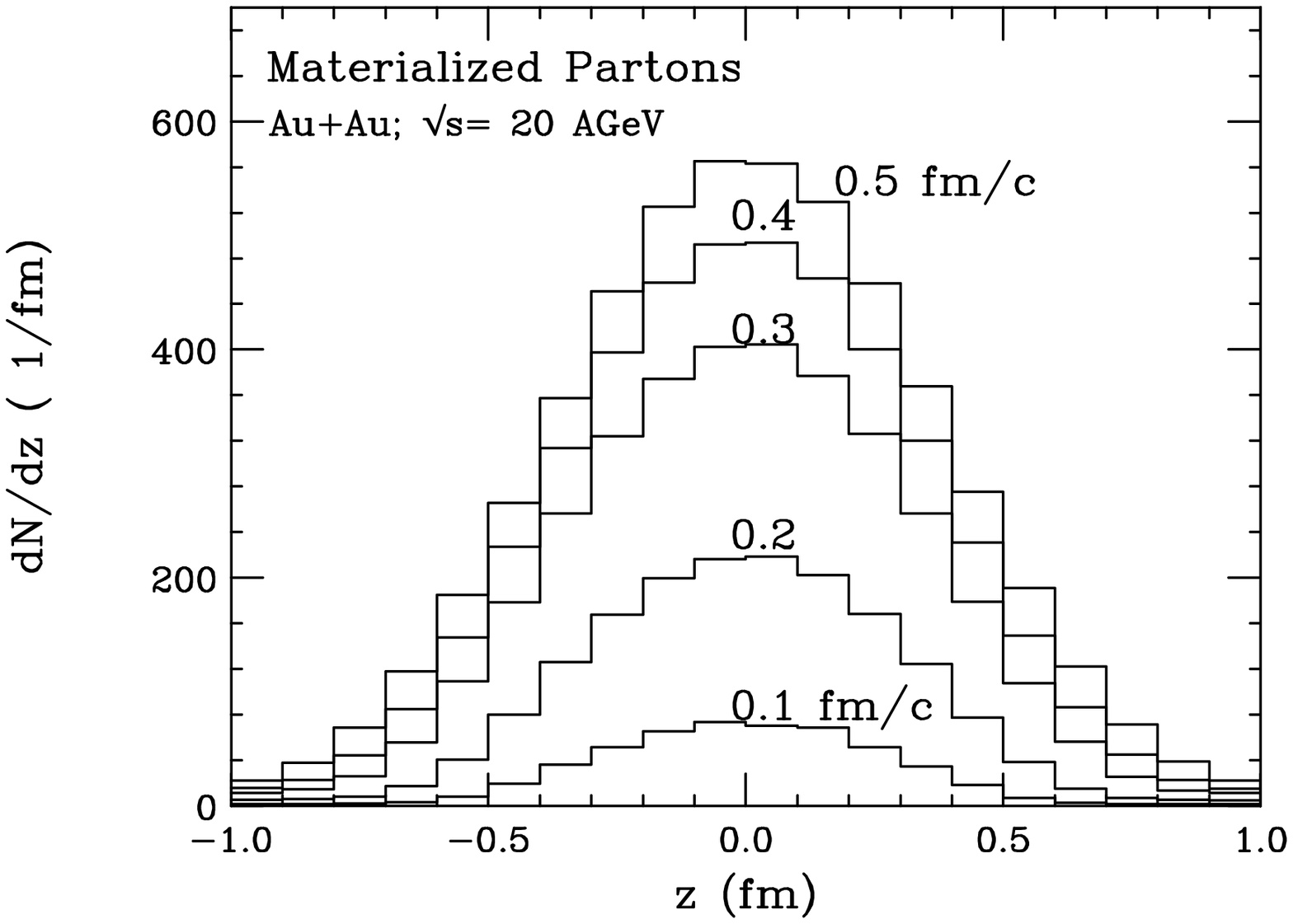} }
\end{figure}
\setcounter{figure}{1}
\begin{figure}
\epsfxsize=250pt
\centerline{ \epsfbox{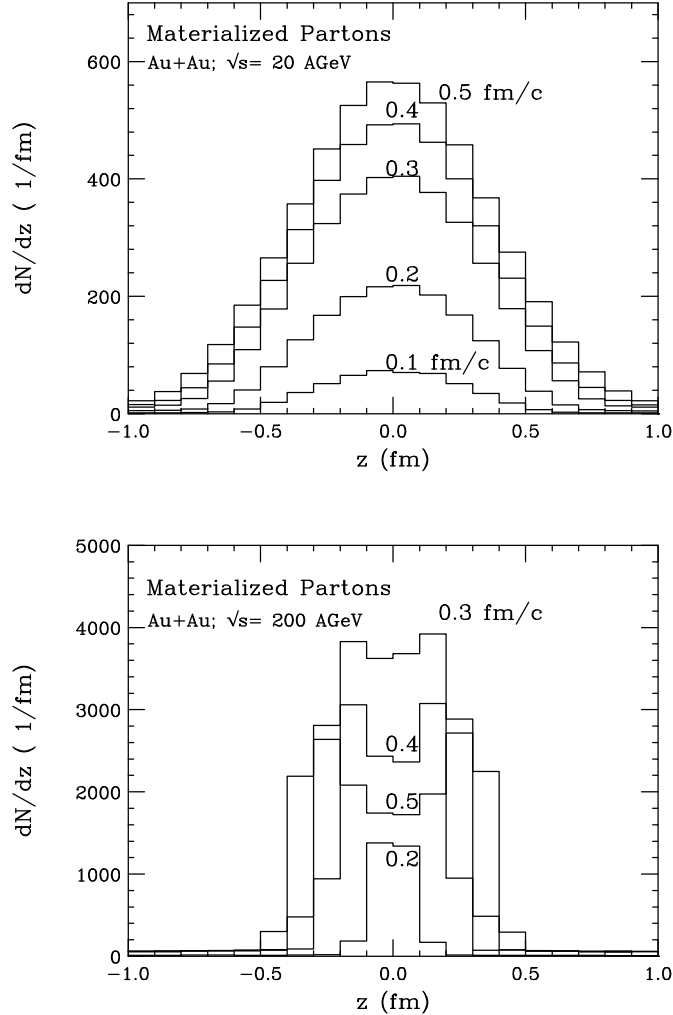} }
\vspace{1.0cm}
\caption{
Distribution of materialized partons per unit length
in the centre of mass frame of the colliding nuclei at different
times along the longitudinal $z$-axis at
 20 A GeV (upper pannel) and 200 A GeV (lower panel).
 The zero of time
is taken as the instant of complete overlap.
}
\end{figure}

In Fig.~2a we plot the number distribution of partons along the
collision axis at several times $0 < t \le 0.5$ $fm/c$ (with
respect to the laboratory frame) for
$\sqrt{s}=$ 20 A GeV, while the corresponding results for 200 A GeV are
given in Fig.~2b. 
 We see that at 20 A GeV the  number density of partons  continues
to increase and peaks at $z=0$ up to a lapse of $t=$ 0.5 $fm/c$. 
Among other things
this is indicative of the fact that until then the crossing over of the
Lorentz contracted ( $\gamma \approx 10$) nuclei is not complete. 
On the other hand the number of partons per unit length reaches its 
maximum around $t=$ 0.3 $fm/c$ at RHIC energy of 200 A GeV. By this time the
the Lorentz contracted ($\gamma \approx 100$) nuclei have crossed each other
and in fact the bulk of the matter is piled away from $z=0$ on either
side.
\smallskip


Figs.~3a and ~3b further support this view through the shape of the
$p_T$ distribution of the partons  at different times.
 We see that the the number of partons having a given $p_T$ increases
rapidly with time at 20 A GeV while at 200 A GeV it reaches a saturation
value by the $t=0.3$ $fm/c$. We also see a noticable bulge in their
number around $p_T \approx p_0$ for the two cases, where $p_0$ is
equal to 1.1 and 2.1 GeV/$c$, respectively.

It is quite instructive to see the time evolution of the number and the
energy density  profile of the partons, shown for the two cases in 
Figs.~4a and 4b for $z=0$.
They attain their maximum values around the common 
time of $\tau=$ 0.3 $fm/c$ for both energies. 
We also find that the number and energy density of the 
partons can get rather large and at RHIC energies it can get up to five
times as large as at SPS energies, around the same point of time.

Finally in Figs.~5a and ~5b we give our predictions for the energy 
density profiles at $z=0$ and $\tau=0.3$ $fm/c$ for the two energies,
for central collisions of several identical nuclei. One interesting 
aspect of these results is that even in
collisions involving sulfur nuclei at SPS energies, 
an energy density $\varepsilon$ above a critical value 
$\varepsilon_{crit} \simeq 2$ GeV/fm$^3$  may  be attained for a short time.
Of course, as expected, 
we get  larger energy densities extending up to larger radii
as we collide heavier nuclei.

\begin{figure}
\epsfxsize=250pt
\centerline{ \epsfbox{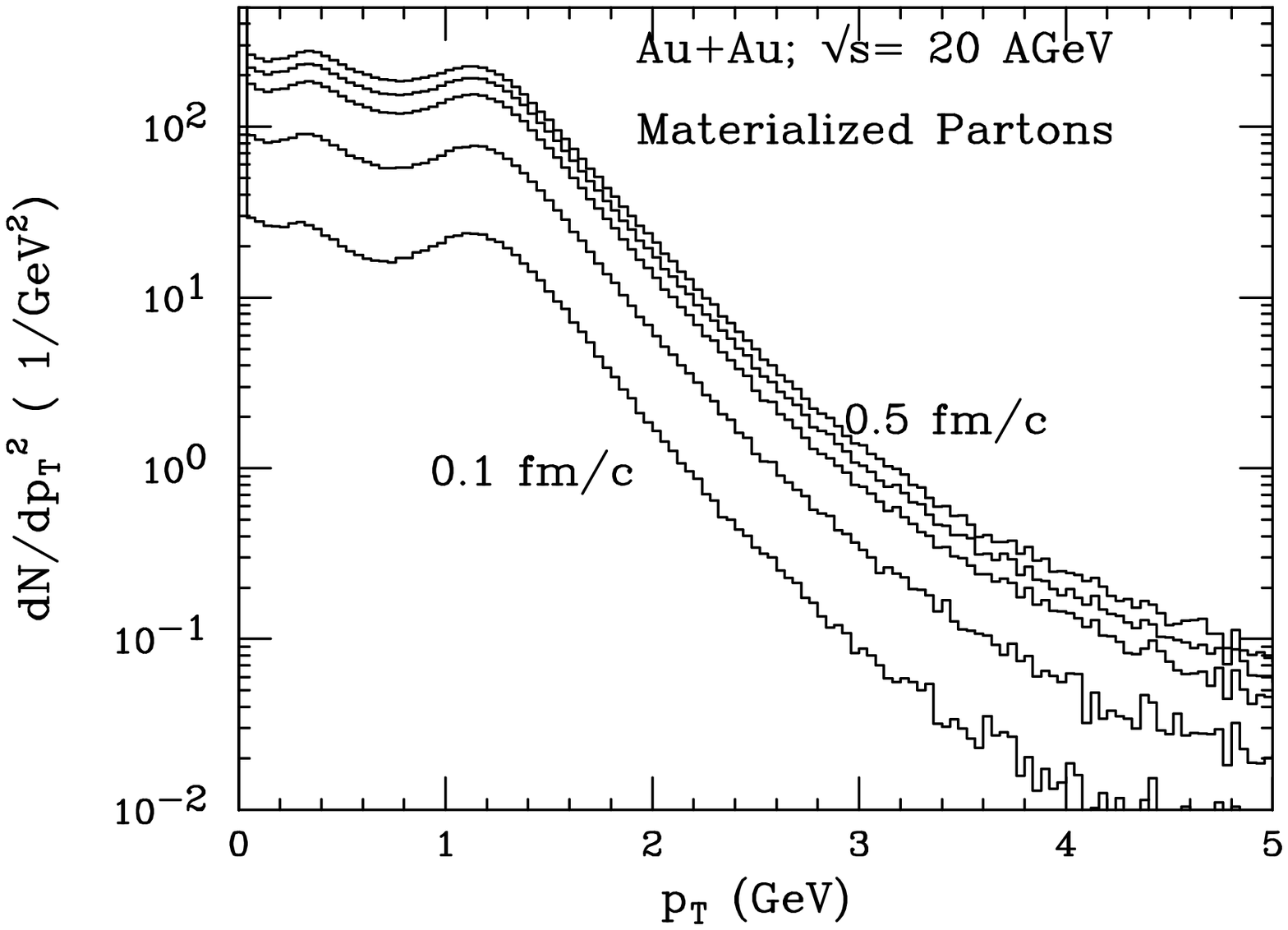} }
\end{figure}
\setcounter{figure}{2}
\begin{figure}
\epsfxsize=250pt
\centerline{ \epsfbox{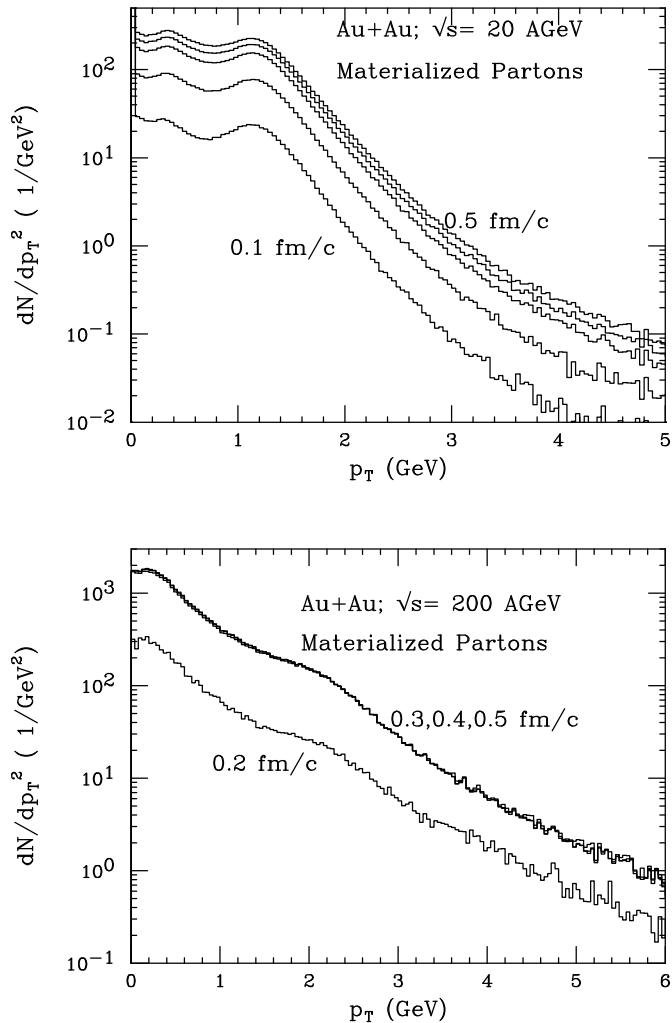} }
\vspace{1.0cm}
\caption{
 The transverse momentum distribution of materialized partons
of the colliding nuclei at different
times for
 20 A GeV (upper pannel) and 200 A GeV (lower panel).
 Only hard scatterings
along with radiations from the partons have been included in the
development of the cascade.
}
\end{figure}

\begin{figure}
\epsfxsize=350pt
\centerline{ \epsfbox{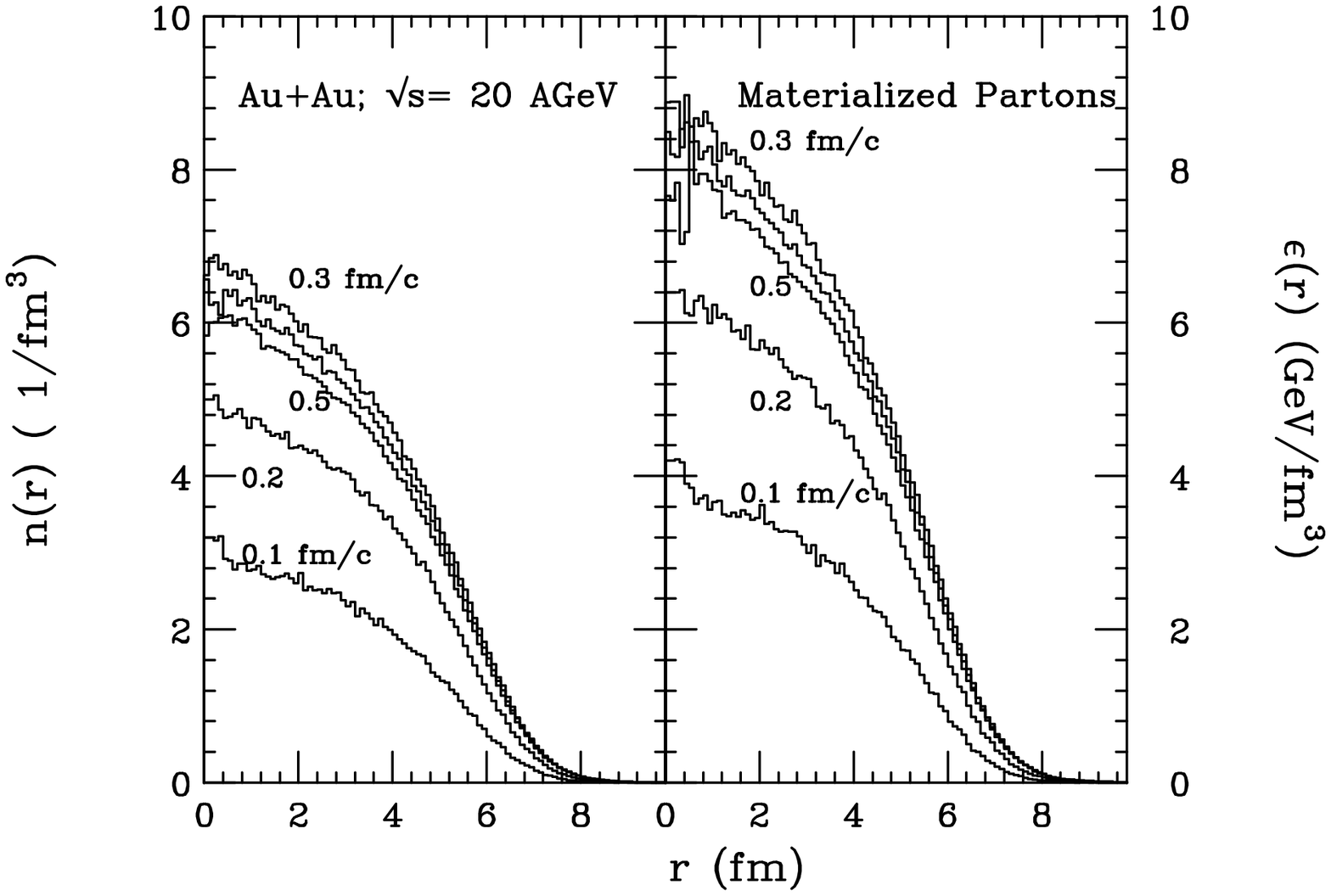} }
\end{figure}
\setcounter{figure}{3}
\begin{figure}
\epsfxsize=350pt
\centerline{ \epsfbox{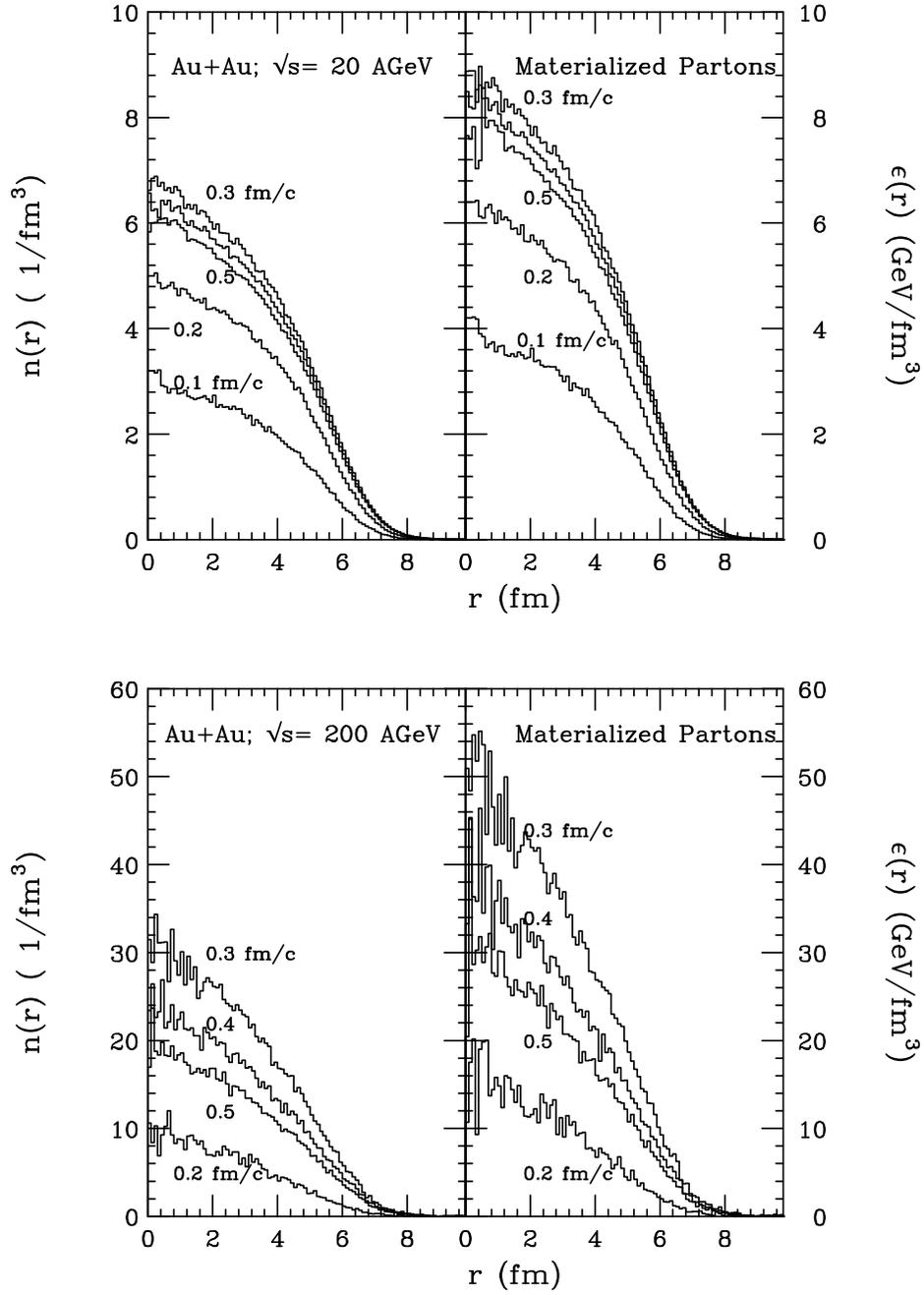} }
\vspace{1.0cm}
\caption{
The time evolution of the number density (left half) and the
energy density (right half) of the materialized partons
in the central region ($!z! \leq$ 0.5 fm)
of the colliding nuclei at
 20 A GeV (upper pannel) and 200 A GeV (lower panel).
 Only hard scatterings
along with radiations from the partons have been included in the
development of the cascade.
}
\end{figure}

\newpage

\begin{figure}
\epsfxsize=250pt
\centerline{ \epsfbox{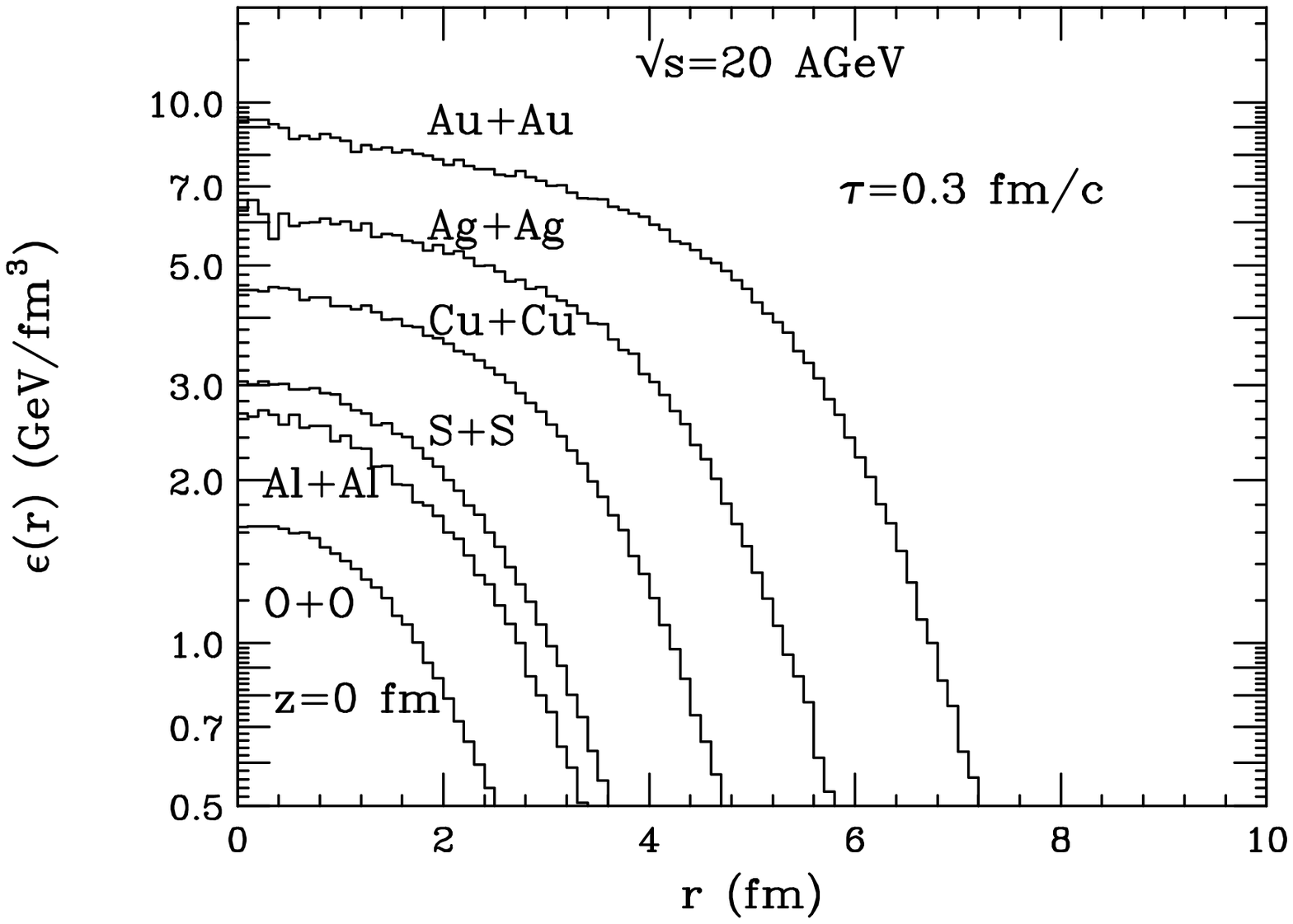} }
\end{figure}
\setcounter{figure}{4}
\begin{figure}
\epsfxsize=250pt
\centerline{ \epsfbox{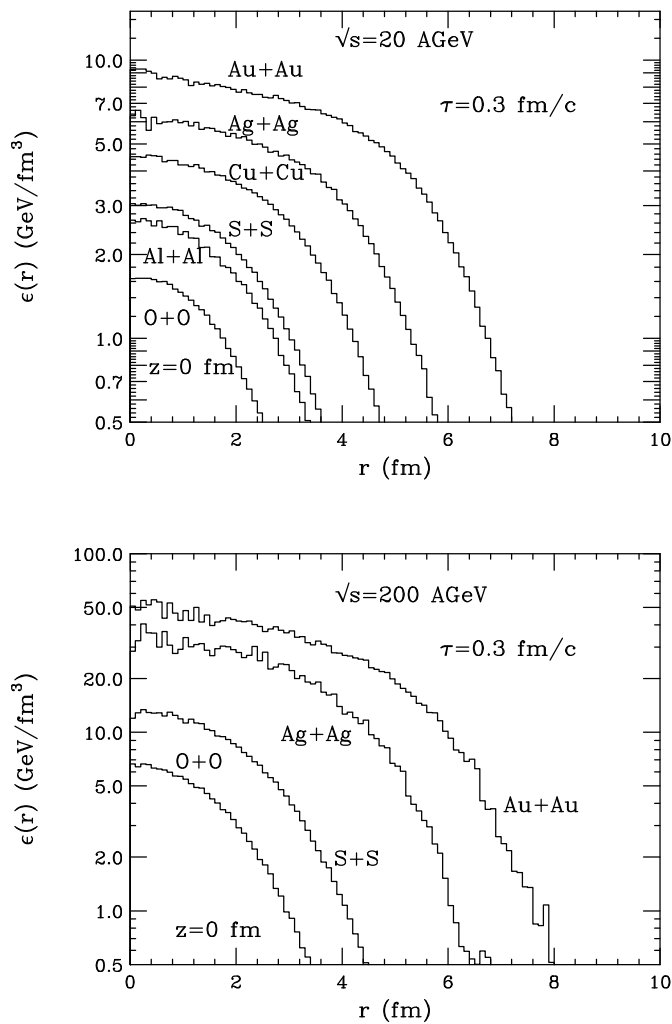} }
\vspace{1.0cm}
\caption{
Energy density of the partonic matter at $\tau=$ 0.3 fm/$c$ for
central collision of identical nuclei at 20 A GeV (upper panel)
and 200 A GeV (ower panel) from parton cascade model.
}
\end{figure}

\subsection{Multiple scattering as response to dense matter}
\smallskip

 Does the creation of a partonic matter which may attain a rather high
energy density also imply enhanced multiple scatterings? Not necessarily,
as a consideration of free-streaming matter readily shows. It is
quite clear that multiple scatterings distinguish a freely streaming
matter from a matter which expands and cools. Thus multiple scatterings
are essential to obtain and maintain a thermal equilibrium, and to
associate a temperature to the system under consideration.
They also facilitate chemical equilibration.
\smallskip

What are the measures of multiple scatterings?
Clearly, if multiple scatterings succeed in maintaining a thermal
equilibrium in an expanding system, then we shall witness a decreasing
temperature with the passage of time.
Moreover, multiple scatterings should increase the total
entropy production, which would result from an enhanced
production of particles. As we have elucidated in the previous
subsection, the  multiplicity of produced
partons reaches a maximum {\it within} a time interval of 0.5 $fm$, subsequent
to which rescatterings may drive the system towards
equilibration.
Therefore one would expect that multiple
parton scatterings manifest themselves in an enhanced production
of gluons, light quarks, and even strange quarks, when compared
to a dynamics with only single, primary parton scatterings as 
in hadronic collisions.
In the latter case the multiplicity would scale {\it linearly}
with the  mass number $A$, whereas in the case of multiple
scatterings, one should see a {\it non-linear} increase of the
multiplicity.
In effect, 
final-state particles that emerge from this early stage, such as  photons, 
dileptons and  strange hadrons, should exhibit such a non-linear
behaviour as one increases the size of the nuclei by going to
heavier collision systems.
 An indication of the emergence of this
aspect was noted in our recent work \cite{dkskkg2}
on the production of single photons
at SPS and RHIC energies.
\medskip

Figs. 6 and 7 exhibit this non-linear behavior 
due to multiple parton scatterings in $A+A$ collisions 
at 20 A GeV and 200 A GeV, respectively, 
in comparison to the same collisions 
accounting only for first scatterings of partons.
In both Figures, the top part shows the increase with $A+A$ of the
total number of produced gluons, light quarks and strange quarks,
while the bottom part shows the partonic rapidity density
$dN^{q+g}/dy|{y=0}$, the transverse energy density
$dE_\perp^{q+g}/dy|_{y=0}$, and the accumulated number 
$N_{hard\,coll}$ of (semi)hard parton-parton collisions.
The symbols are the results of our simulations for $A+A$
from Helium all the way up to Uranium, whereas the lines 
serve to guide the eye.
\smallskip

\begin{figure}
\epsfxsize=250pt
\centerline{ \epsfbox{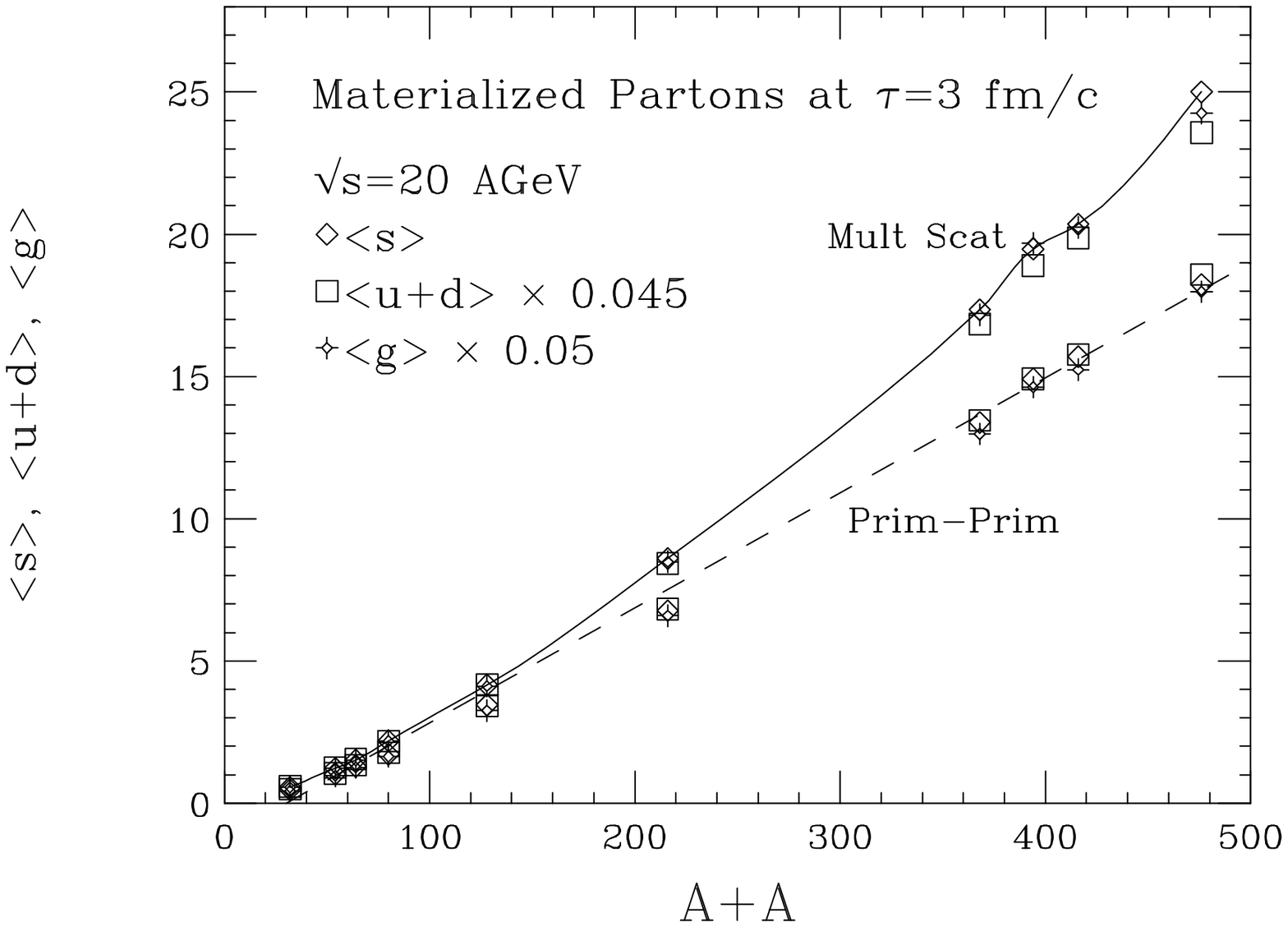} }
\end{figure}
\setcounter{figure}{5}
\begin{figure}
\epsfxsize=250pt
\centerline{ \epsfbox{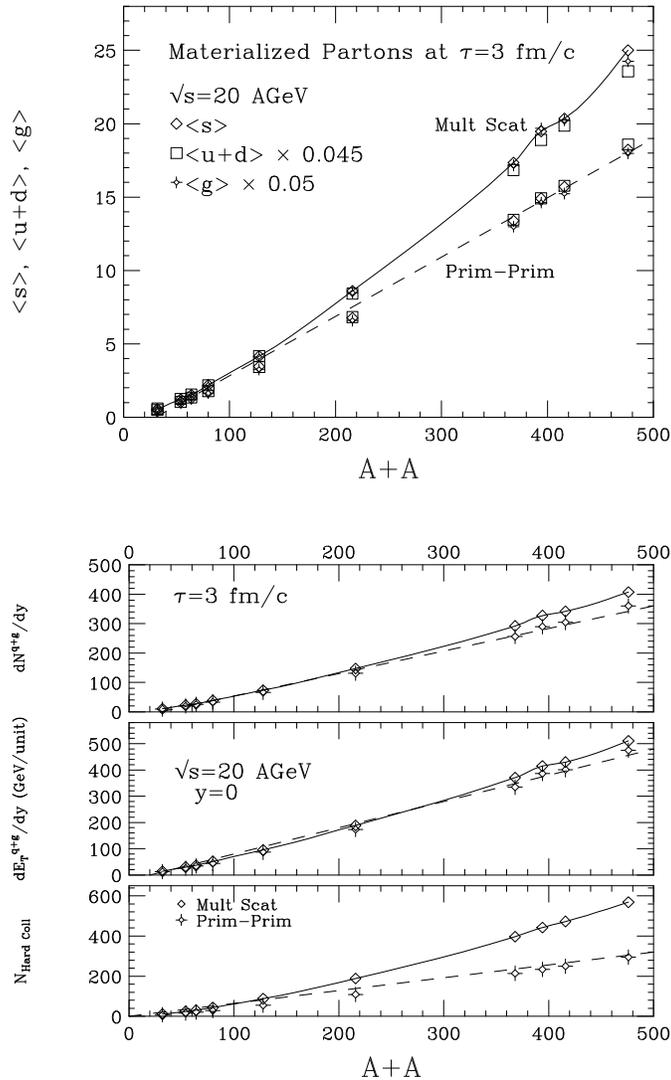} }
\vspace{1.0cm}
\caption{
Effect of multiple scatterings on
the variation of number of materialized partons at $\tau=$ 3 fm/$c$
as a function of number of participating nucleons in central
collision of identical nuclei at $\sqrt{s}=$ 20 A GeV (uppermost panel).
The three lower panels show the effect of multiple scatterings
on the total number of hard collisions, and the rapidity
density of partons and their transverse energy.
}
\end{figure}
\bigskip

The common feature of all the plots is the non-linear   
departure of the multiple scattering picture (solid lines) from a  linear
scaling with $A$ in the case of only primary scatterings (dashed lines),
which sets in for systems $A+A$ above 220, corresponding to
collisions of Silver nuclei.
Also remarkable is that the qualitative behavior of this non-linearity is
the same for $\sqrt{s}=20$ A GeV (Fig. 6) and for
$\sqrt{s}=200$ A GeV (Fig. 7), however, the absolute magnitude
of the various quantities shown is of course drastically different
due to the factor 10 difference in the total energy.
For example, the multiplicities of materialized gluons,
light and strange quarks at 200 A GeV are about a factor of six larger
for the heaviest system, than the corresponding numbers for 20 A GeV.
Most remarkable, however, is the substantial difference
betewen the multiple and the single scattering scenarios:
at $\sqrt{s}=20$ A GeV, the multiple scatterings enhance
the materialized parton numbers up to 25-30 \%, whereas
at $\sqrt{s}=200$ A GeV, the amplification can reach up to 40 \%.

\begin{figure}
\epsfxsize=250pt
\centerline{ \epsfbox{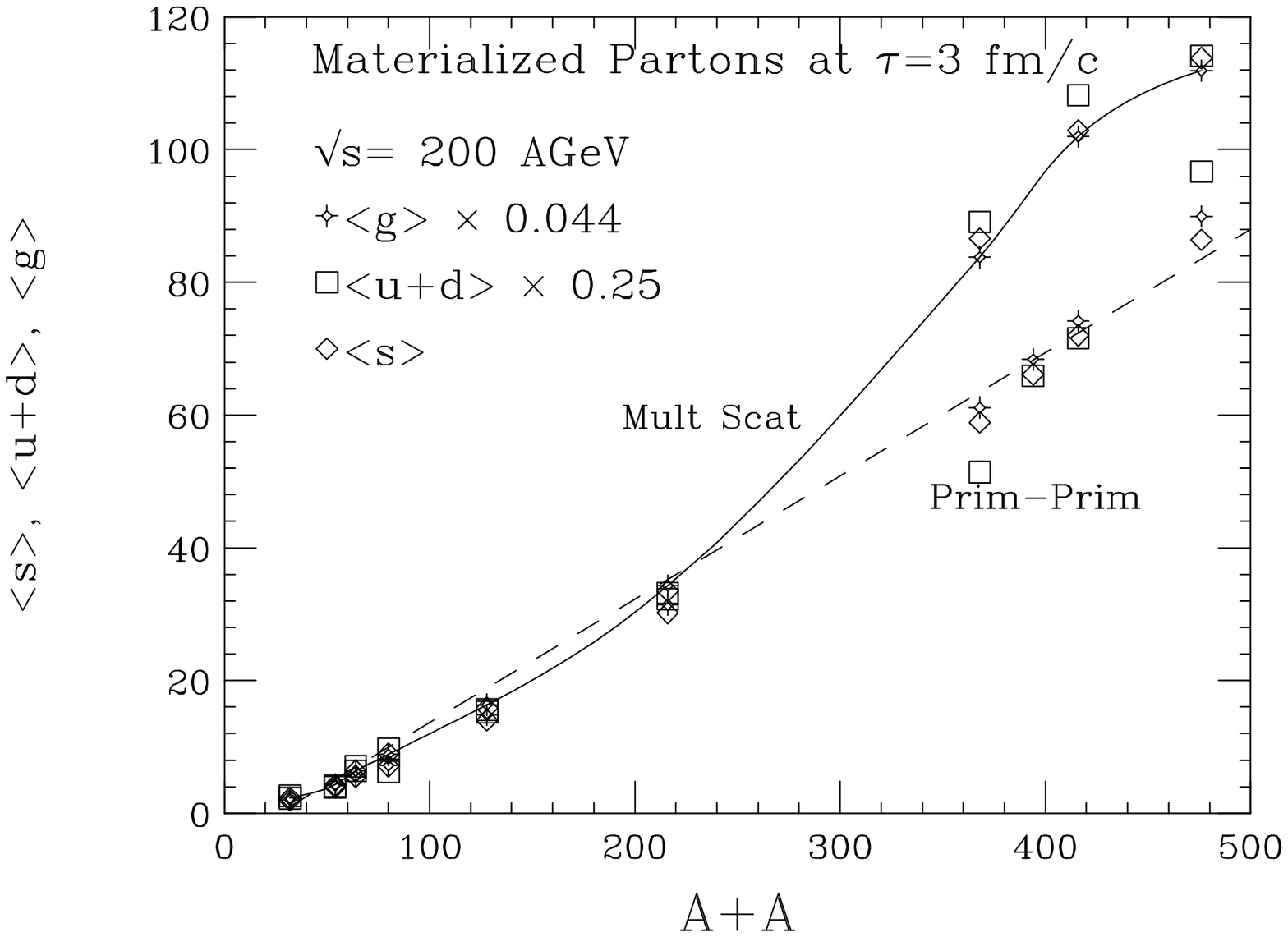} }
\end{figure}
\setcounter{figure}{6}
\begin{figure}
\epsfxsize=250pt
\centerline{ \epsfbox{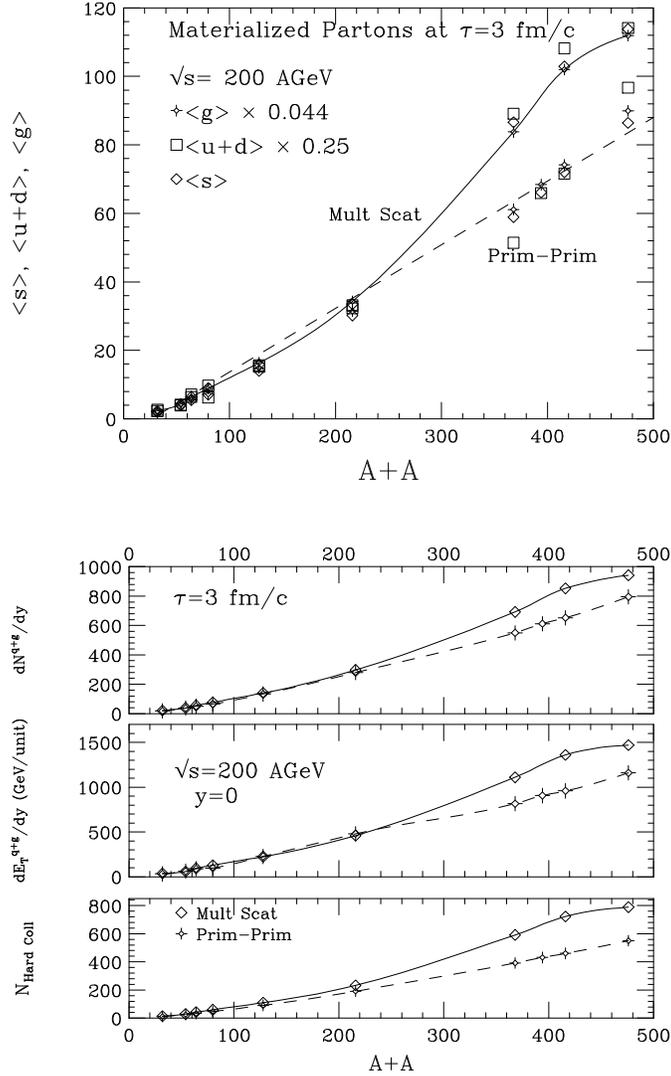} }
\vspace{1.0cm}
\caption{
Effect of multiple scatterings on
the variation of number of materialized partons at $\tau=$ 3 fm/$c$
as a function of number of participating nucleons in central
collision of identical nuclei at $\sqrt{s}=$ 200 A GeV (uppermost panel).
The three lower panels show the effect of multiple scatterings
on the total number of hard collisions, and the rapidity
density of partons and their transverse energy.
}
\end{figure}

\newpage
\subsection{Interplay between multiple scatterings and parton emissions}
\smallskip

So far we have discussed the role of multiple scatterings alone,
and have not addressed the relevance of  gluon
and quark-antiquark production through multiple emission processes.
As we have mentioned in the Introduction,
one cannot really disentangle the 
parton emission processes from the scatterings, because the latter
are the source of the former, but the former in effect influence the
latter:
Each (semi)hard parton-parton scattering is a potential trigger
for a bremsstrahlung cascade of emissions, if the scattering
was sufficiently inelastic.
Therefore the emission of partons stimulated by scatterings
grows naturally when the number of parton collisions increases.
On the other hand, an enhanced emission of quanta feeds back on
the dynamics by increasing the local parton density and hence
the rescattering probability as well as the probability
for further emissions.
\smallskip

What limits this avalanche effect and prevents a parton population
catastrophe?
There are three mechansims: The first one is the multi-particle
dynamics in heavy-ion collisions itself, which eventually causes the
system to expand and diffuse rapidly, and hence the
interactions cease.
The second is due to energy conservation, which limits
the maximum amount of energy that can be harnessed in the particle 
interactions, and results in lower-energetic interactions 
with decreasingly smaller cross-sections as time goes on.
The third mechanism is the Landau-Pomeratchuk-Migdal effect \cite{LPM1,LPM2},
which is mimicked in the PCM by assigning each parton
that emerges from a collision or an emission, a specific
formation time $\tau = E/q_\perp^2$ during which the parton
cannot interact. As a consequence, the production and the re-interaction of
particles is the more delayed the smaller the momentum transfer
$q_\perp$ of the interaction is,
which  causes a limitation on the parton density
increase, because it is mainly due to the production
of low-energy gluons with the largest re-interaction
cross-sections.
The elongated formation time for these soft gluons
makes time for the dense region to expand and dilute.

\smallskip
It is clear that  the interplay between collisions and stimulated emission
is a complex net of interactions which is difficult to disentangle.
Nevertheless, by switching on/off multiple scattering and
QCD radiation, one may get at least a qualitative impression
of the different parton cascade aspects.
In Fig. 8 we plot the results for three different scenarios:
The first scenario (dashed-dotted lines)
 includes solely elastic scatterings
with all parton emission switched off,
and moreover considers only single collisions of 
initial-state partons, i.e., the first scatterings of two
primary partons with no further collision of those.
The second scenario (dashed lines) 
includes now inelastic collisions
with radiative emission of partons,  but still allows only
for single collisions of primary partons.
Finally, the third option (solid lines) is the full parton-cascade 
development with multiple parton scatterings plus associated parton decays.
Fig. 8a (top) shows for $Au+Au$ collisions at $\sqrt{s}= 20$ A GeV
the rapidity spectra $dN/dy$ of materialized quarks and gluons
(left panel) and the corresponding transverse energy distributions
$dE_\perp/dy$ at $t = 3$ $fm/c$. Fig. 8b (bottom)
displays the same quantities for $\sqrt{s}= 200$ A GeV at
$t = 3$ $fm/c$.
\smallskip 

The qualitative 
features of the $dN/dy$ spectra, common to both energies, are a
substantial contribution to particle production
from radiative emissions. For example, even if only
primary collisions are considered, the associated parton
emission increases $dN_/dy$ at $y=0$ by about a factor of
1.5 at 20 A GeV and by more than a factor of 2 at 200 A GeV,
as compared to primary collisions with radiative emission switched off.
Allowing then also for multiple scatterings, one observes a further
particle enhancement at mid-rapidity, however much smaller than
the difference due to parton emission.
These results are immediately plausible, because 
primary scatterings involve, on the average, the most energetic 
partons with their full initial energy, and therefore
most primary scatterings are inelastic, accompanied by the emission
of at least one gluon. On the other hand, secondary scatterings 
involve mainly  partons that have lost already energy in previous
interactions, so that the majority of them is of elastic nature
without gluon emission.
Therefore primary collisions provide a substantial parton production
from emission processes, whereas secondary rescatterings do
not drastically change the multiplicity further, but mainly
redistribute energy and momentum among the particles.
Turning to the $dE_\perp/dy$ spectra, the growth of the transverse
energy at $y=0$ from primary collisions without and with radiation to
multiple scatterings with radiation, follows the rapidity distributions
just discussed.
However, the differences between the three curves are much less
pronounced. This is due to the fact that a large part of the multiplicity
arises from low-energy radiative gluons, which leave a prominent
mark in the multiplcity spectra, but add little to the
total transverse energy.

\smallskip

\newpage

\begin{figure}
\epsfxsize=350pt
\centerline{ \epsfbox{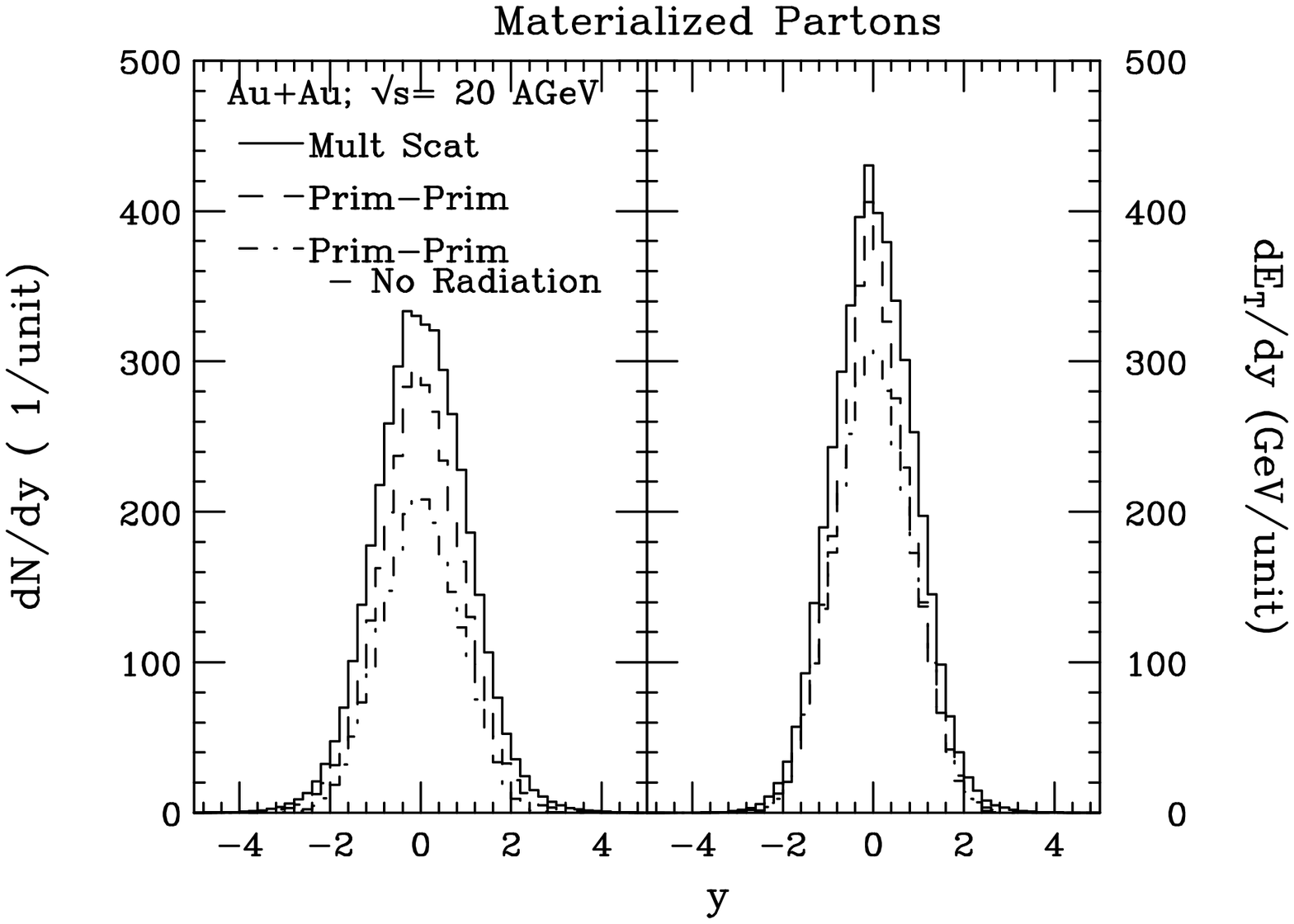} }
\end{figure}
\setcounter{figure}{7}
\begin{figure}
\epsfxsize=350pt
\centerline{ \epsfbox{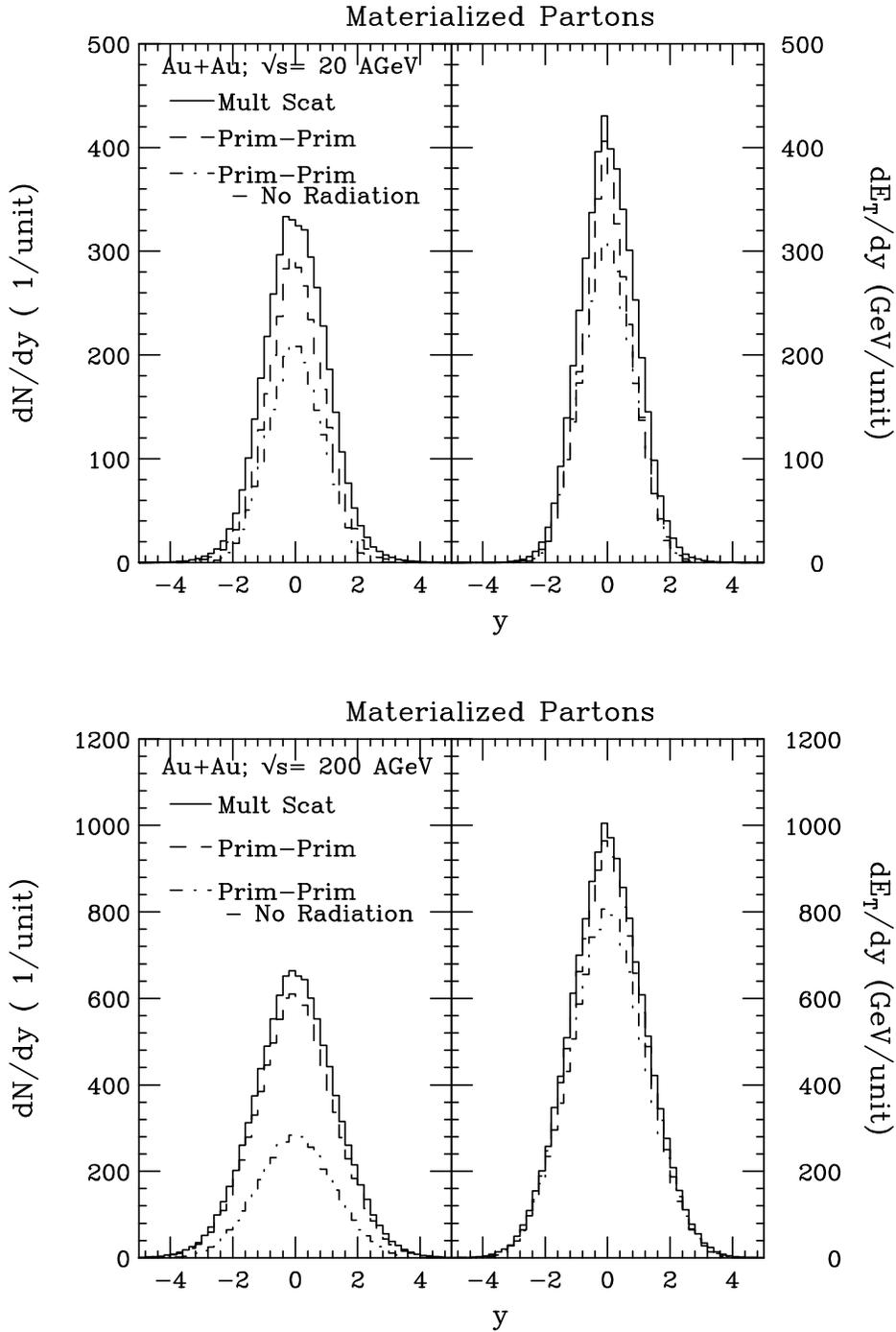} }
\vspace{1.0cm}
\caption{
Consequenses of radiations and multiple scatterings on
rapidity density of number of partons (left half) and
their transverse energy (right half) at  $\sqrt{s}=$ 20 A GeV
(upper panel) and 200 A GeV (lower panel). Only hard
scatterings are included and the parton cascade is evolved till
$t=$ 3 fm/$c$.
}
\end{figure}

In Fig. 9, the $p_T$ distribution the partons in the central rapidity region
is seen to also reflect many of these developments in an interesting manner.
It is evident that the simulations with only primary - primary collsions
lead to a distribution of partons with a noticable peak around
$p_T\approx p_0$ (where $p_0 = 1.1$ (2.1) GeV/$c$ for 
$\sqrt{s} = 20$ (200) A GeV).
The radiations of gluons then reallocates the
high $p_T$ partons to low $p_T$ region, and their number also
goes up considerably. The multiple scatterings, however,
affect the $p_T$ spectra differently at 20 A GeV and 200 A GeV.

The  most interesting aspect is that the
multiple scatterings enhance the number of partons having large
$p_T$ at 20 A GeV, while they decrease their number at 200 A GeV.
The origin lies in the fact that once the scatterings
lead to partons which do not have large virtualities, they would not
radiate, and thus the scatterings will lead to increase in $p_T$.
This is appearently the case at 20 A GeV, wheras 
at 200 A GeV the virtualities of the scattered partons 
could still be large and lead to radiation of gluons having
low $p_T$, and thus the multiple scatterings do not enhance the
$p_T$.

\begin{figure}[htb]
\begin{minipage}[t]{80mm}
\epsfxsize=200pt
\rightline{ \epsfbox{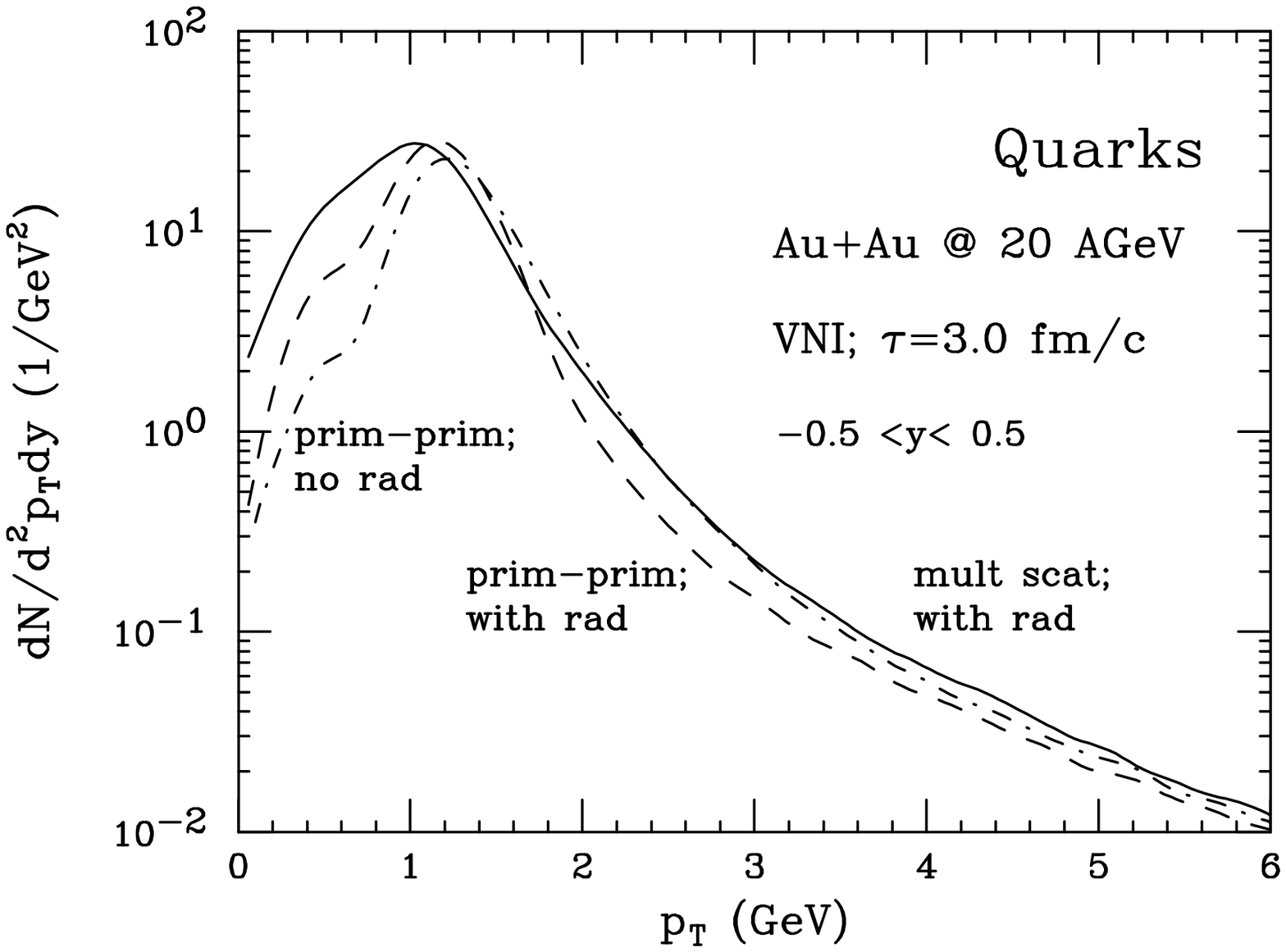} }
\vspace{-0.8cm}
\end{minipage}
\hspace{\fill}
\begin{minipage}[t]{75mm}
\epsfxsize=200pt
\centerline{ \epsfbox{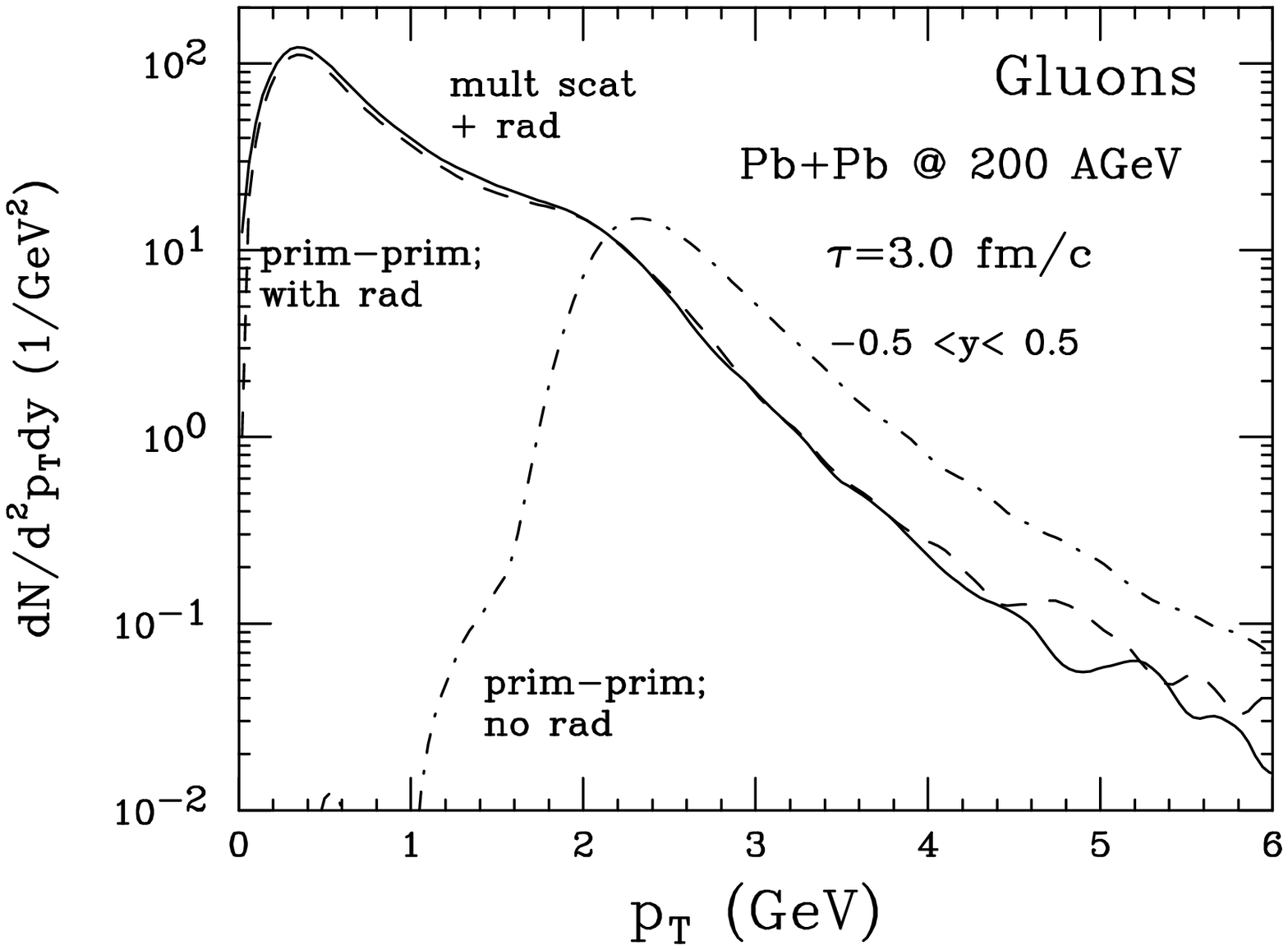}  }
\vspace{-0.8cm}
\end{minipage}
\begin{minipage}[t]{83mm}
\epsfxsize=200pt
\rightline{  \epsfbox{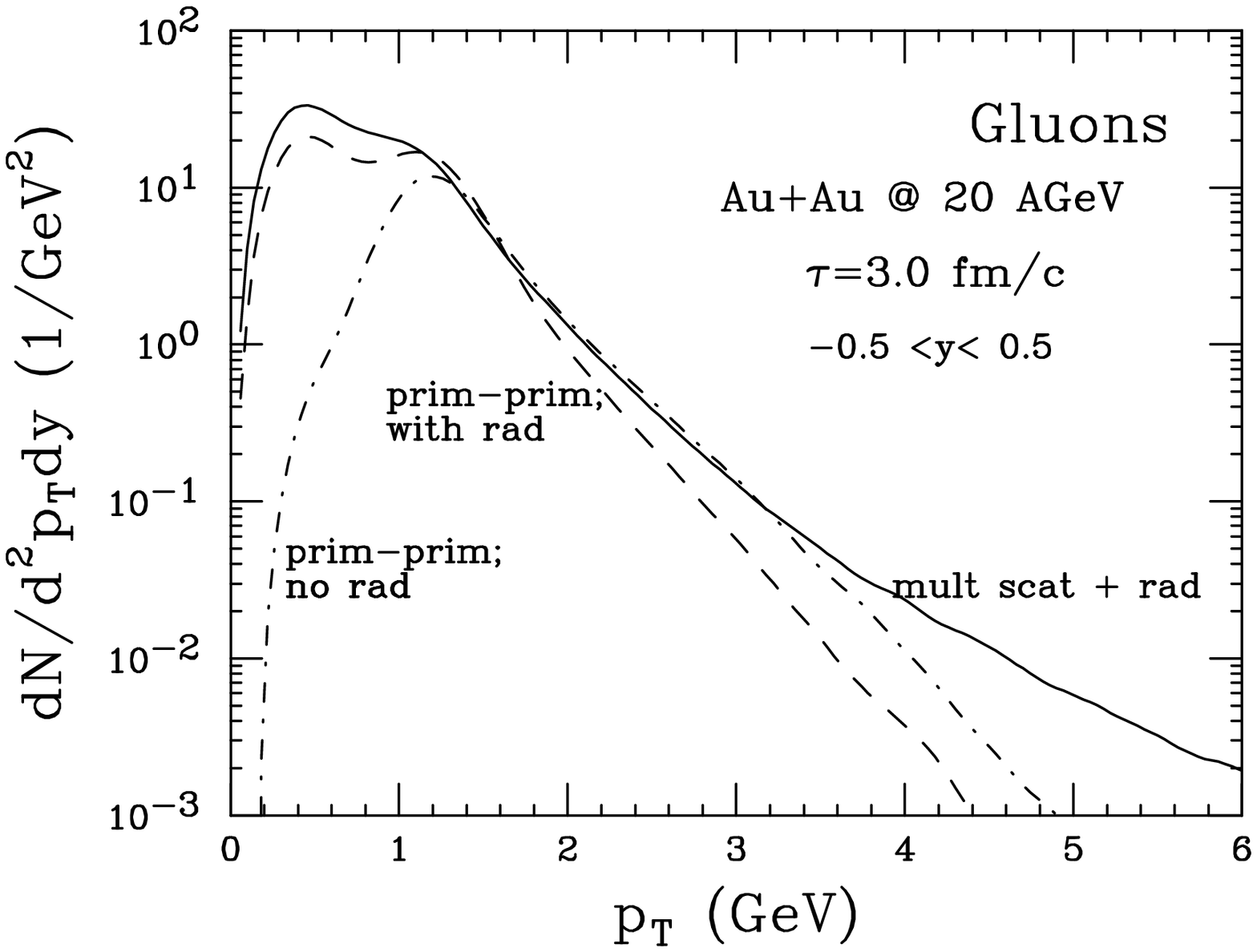} }
\vspace{-0.8cm}
\end{minipage}
\hspace{\fill}
\begin{minipage}[t]{75mm}
\epsfxsize=200pt
\centerline{ \epsfbox{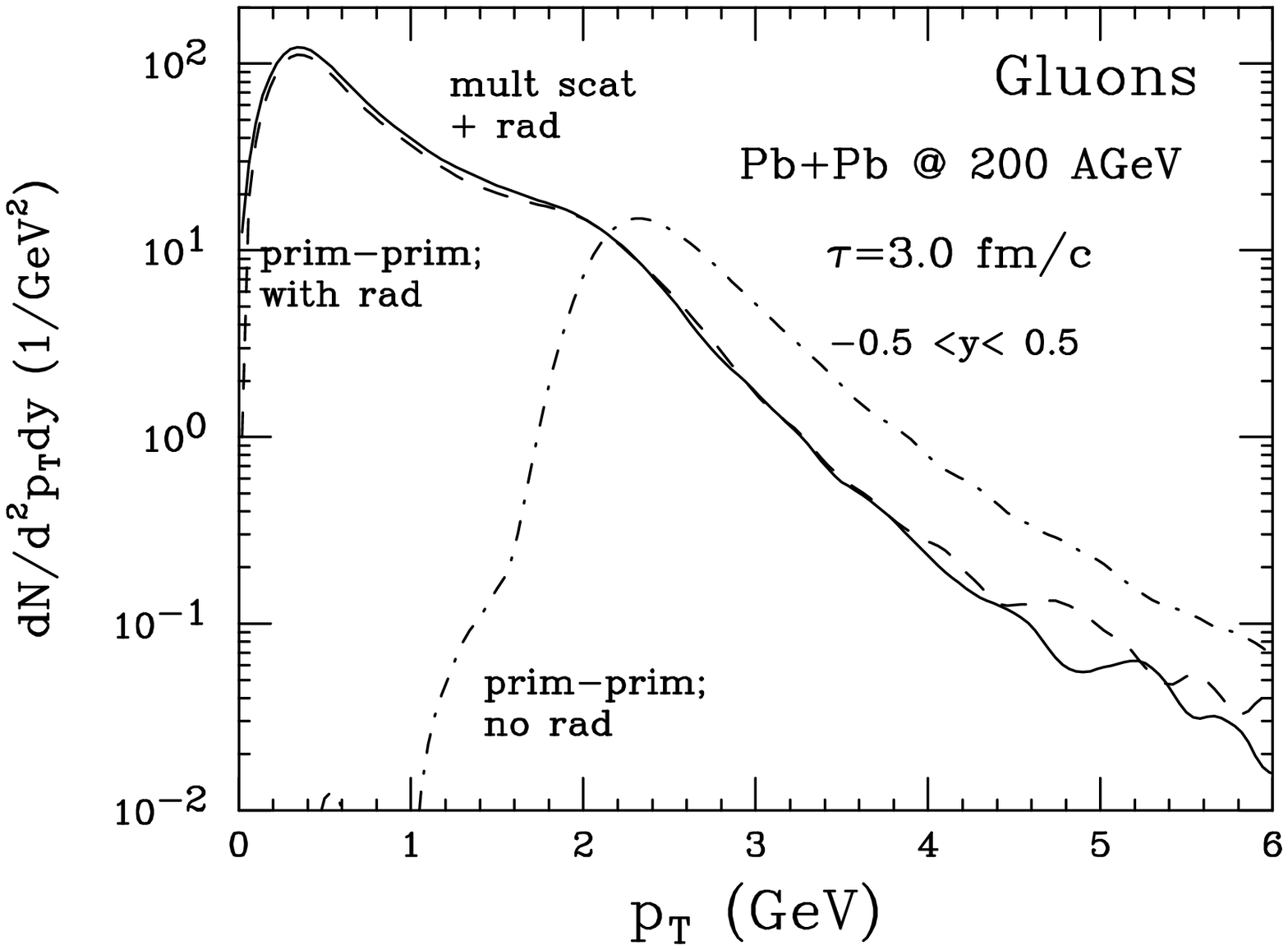}  }
\vspace{1.0cm}
\end{minipage}
\caption{ 
Interplay of radiations and multiple scatterings on
the transverse momentum distribution of quarks and 
gluons in the central rapidity region at
$\tau=$ 3 fm/$c$ at $\sqrt{s}=$ 20 A GeV (left panels) and 200 A Gev
(right panels).
Only hard scatterings leading to $p_T >p_T^0$, where $p_T^0 \approx $ 1.1
GeV at 20A GeV and $\approx 2.1$ GeV at 200 A GeV, are included.
          }
\end{figure}

\bigskip

\section{Summary}
Within the PCM framework and using the computer simulation 
program VNI,
we have studied the development of dense partonic
systems in central collsions of several nulcei at 20 A GeV
and 200 A GeV, in order to bring out the role of 
quark-gluon cascades due to multiple
scatterings and radiative emission of partons during the
early stage of these collisions.
Since we restricted our analysis to the perturbative QCD sector
and avoided confronting model-dependencies associated
with a hadronization prescription, we are confident
that our results draw a fairly realistic picture
of the short-range QCD dynamics at high density.
Our main findings may be summarized as follows:

\begin{description}
\item[(i)]
The initial materialization of
partons through scatterings or radiative emission reaches
its maximum around 0.5 $fm/c$ (at 20 A GeV)
and 0.3 $fm/c$ (at 200 A GeV).
After that lapse, the liberation of initial-state partons
damps out strongly, however, radiative parton emissions  
and rescatterings between these produced partons continue.
\item[(ii)]
The total number of produced partons around $z=0$ achieves its maximum
of about 600 for 20 A GeV and close to 4000 at 200 A GeV for
central collision of gold nuclei. This
increase of a factor 6.5 reflects itself in the corresponding 
number densites which increase by a factor of 5 from
20 A GeV to 200 A GeV, and an even stronger increase
in the energy density at $z=0$ from $\sim$ 8 GeV/fm$^3$ to
to $\sim$ 50 GeV/fm$^3$.
\footnote{
Note that due to the different Lorentz
contraction and the different times (0.5 $fm/c$ and 0.3 $fm/c$)
at which the maximum parton multiplicity is achieved,
the associated volume of the collision system is not the
same in the two cases.}
\item[(iii)]
Multiple scatterings along with the associated parton emission
lead to a non-linear increase with nuclear mass number $A$
of parton multiplicity and number of parton collisions,
as compared to single, primary scatterings only, in
which case the growth pattern scales linearly with $A$.
\item[(iv)]
The emission of partons in radiative processes associated 
parton collisions contributes substantially to the multiplcity,
making out up to 30 \% at 20 A GeV and about 50 \% at 200 A GeV.
The radiative emissions are mainly due to
inelastic, primary scatterings of initial-state partons,
whereas secondary scatterings are to large extent of elastic
nature without additional emission, 
but do redistribute energy and momentum among the particles.
\end{description}
\medskip

What should be done in the near future, is to establish a
connection between the above partonic properties and
the final-state hadronic and electromagnetic particles.
With a suitable parton-hadron conversion model, 
one must attempt to identify the high-density
aspects of parton cascades in measurable signatures that
one can extract from data. For instance:
\begin{itemize}
\item
Does the the final-state transverse energy
spectrum $\frac{dE_\perp}{dy}$ reflect the non-linear behaviour
of the corresponding partonic quantity? How
does, e.g., the ratio 
$\frac{dE_\perp}{dy}(200 AGeV)/\frac{dE_\perp}{dy}(20 A GeV)$
scale?
How does the event distribution distribution $\frac{dN_{event}}{dE_\perp}$ 
of the total transverse energy per event (which has the famous 
`horseback' shape) stretch out to larger $E_\perp$ as the
collision system or the collision energy is increased?
\item
Can one actually `see' some  remnants of minijets and 
parton cascades in final-state observables? E.g., how
does the enhancement of the angular correlation (between 
pairs of partons) at small angles due to evolving cascades
 show up in the hadron pair-correlations?
In the absence of minijets, one would expect a flat
correlator when plotted versus the relative angle, while
for prominent and energetic cascading one would
see a pronounced peak at very small angles.
\end{itemize}
This incomplete list gives examples of experimental
quantities, which are presumably difficult to
trace back to the partonic history
due to insufficient statistics, but that can 
already be analysed at the end of
{\sl day one} at BNL RHIC when it starts operation.
\bigskip
\bigskip
\bigskip


\section*{ACKNOWLEDGEMENTS}

This work was supported in part by the D.O.E. under contract no.
DE-AC02-76H00016.


\bigskip
\bigskip
\bigskip

\newpage

\begin{thebibliography}{39}


\bibitem{scott} For a general theoretical overview, see:
                W. T. Scott, Rev. Mod. Phys. {\bf 35}, 231 (1963).

\bibitem{msrep} K. Geiger, Phys. Rep. {\bf 258}, 376 (1995).

\bibitem{gyulassywang} M. Gyulassy and X.-N. Wang, 
                       Nucl. Phys. {\bf B420}, 583 (1994).

\bibitem{BDMPS} R. Baier, Yu.L. Dokshitser, A.H. Mueller, S. Peigne, 
                and D. Schiff,
                Phys. Lett. {\bf B345}, 277 (1995); 
                Nucl. Phys. {\bf B483}, 291 (1997); 
                Nucl. Phys. {\bf B484}, 265 (1997). 


\bibitem{qiu}   J.-W. Qiu and G. Sterman, in the
                {\it Proceedings of the RHIC Summer Study 96},
                Brookhaven National Laboratory, 8-19 July 1996,
                $\langle$ hep-ph/9610476 $\rangle$. 


\bibitem{pcm}   K. Geiger and B. M\"{u}ller, Nucl. Phys. B{\bf 369}, 600 (1992);
		K. Geiger, Phys. Rev. D {\bf 47}, 133 (1993).

\bibitem{ms3942} K. Geiger, Phys. Rev. D{\bf 54}, 949 (1996);
                 Phys. Rev. D{\bf 56}, 2665 (1997).

\bibitem{EG}  J. Ellis and K. Geiger, Phys. Rev. D{\bf 52}, 1500 (1995); 
              Phys. Rev. D {\bf 54}, 1967 (1996);
              J. Ellis, K. Geiger, and H. Kowalsky, Phys. Rev. Phys. Rev.
              D{\bf 54}, 5443 (1996).

\bibitem{vni} K. Geiger,  Comp. Phys. Com. {\bf 104}, 70 (1997).
	      The latest version of the computer program VNI can be obtained 
	      from {\it http://rhic.phys.columbia.edu/rhic/vni}, or from the
	      authors.

\bibitem{dkskkg1} K. Geiger and D. K. Srivastava, Phys. Rev. C{\bf 56}, 2718 (1997);
	     D. K. Srivastava and K. Geiger, 
             Phys. Lett. {\bf B422}, 39 (1998);
	     K. Geiger, $\langle$ nucl-th/9801007 $\rangle$, to appear in 
	     Nucl. Phys. {\bf A}.

\bibitem{dkskkg2} D. K. Srivastava and K. Geiger, 
	     $\langle$ nucl-th/9802034 $\rangle$, submitted to 
	     Phys. Rev. {\bf C}.

\bibitem{LPM1}  L. D. Landau and I. Ya. Pomeranchuk,
               Dokl. Akad. Nauk SSSR {\bf 92} ,535 (1953); {\it ibid.}, 735;
               A. B. Migdal,
               Dokl. Akad. Nauk SSSR {\bf 96}, 49 (1954).

\bibitem{LPM2}  N. N. Nikolaev,  Sov. Phys. Usp. {\bf 24}, 531 (1981);
               F. Niedermeyer,  Phys. Rev. {\bf D34}, 3494 (1986). 
	       Rev.  Mod.  Phys. {\bf 56}, 579  (1984).


\end{thebibliography}
\end{document}